%% file: paper.tex
\documentclass[twocolumn,preprintnumbers,superscriptaddress,aps,nofootinbib]{revtex4-2}
\usepackage{amsmath,amssymb,mathtools}
\usepackage{graphicx}
\usepackage{booktabs,cellspace}
\usepackage{color}
\usepackage{hyperref}
\usepackage{xspace}
\usepackage{mathtools}
\usepackage[normalem]{ulem}
\usepackage{soul}
\usepackage{mathrsfs}
\usepackage{subcaption}
\makeatletter 
\renewcommand\onecolumngrid{% <<<<<<
\do@columngrid{one}{\@ne}%
\def\set@footnotewidth{\onecolumngrid}% <<<<<<<<<<<<<<<<
\def\footnoterule{\kern-6pt\hrule width 1.5in\kern6pt}%
}
\makeatother
\newcommand{\GeV}{\ensuremath{\,\mathrm{GeV}}\xspace}

\definecolor{darkgreen}{rgb}{0,0.4,0}
\definecolor{grey}{rgb}{0.5,0.5,0.5}
\definecolor{orange}{rgb}{0.9,0.5,0.0}
\definecolor{lightblue}{rgb}{0.0,0.5,1.0}
\usepackage[dvipsnames]{xcolor}
\usepackage{pdfcomment}
\newcommand{\comment}[1]{#1}

\newcommand{\PKUaff}{School of Physics, Peking University, Beijing,
  100871, China}
\newcommand{\CHEPaff}{Center for High Energy Physics,
 Peking University, Beijing 100871, China}
\newcommand{\ZUaff}{Zhejiang Institute of Modern Physics, School of
 Physics, Zhejiang University, Hangzhou, Zhejiang, 310027, China}
\newcommand{\CERNaff}{CERN, Theoretical Physics Department, CH-1211
 Geneva 23, Switzerland}

\begin{document}

\title{Scaling violation in power corrections to energy correlators
  from the light-ray OPE}

\preprint{CERN-TH-2024-084}

\author{Hao Chen} \affiliation{\ZUaff}%
\author{Pier Francesco Monni} \affiliation{\CERNaff}  %
\author{Zhen Xu} \affiliation{\ZUaff}%
\author{Hua Xing Zhu} \affiliation{\PKUaff}\affiliation{\CHEPaff}

\begin{abstract}
  In recent years, energy correlators have emerged as a powerful tool
  to explore the field theoretic structure of strong interactions at
  particle colliders.
  In this Letter we initiate a novel study of the non-perturbative
  power corrections to the projected $N$-point energy correlators in
  the limit where the angle between the detectors is small.
  Using the light-ray operator product expansion (OPE) as a guiding
  principle, we derive the power corrections in terms of two
  non-perturbative quantities describing the fragmentation of quarks
  and gluons.
  In analogy with their perturbative leading-power counterpart, we
  show that power corrections obey a classical scaling behavior that
  is violated at the quantum level. This crucially results in a
  dependence on the hard scale $Q$ of the problem that is calculable
  in perturbation theory.
  Our analytic predictions are successfully tested against Monte Carlo
  simulations for both lepton and hadron colliders, marking a
  significant step forward in the understanding of these observables.
\end{abstract}

% \pacs{12.38.-t}
\maketitle

\paragraph*{Introduction.---}
Within the physics programme of the Large Hadron Collider (LHC),
energy correlators~\cite{Basham:1978bw,Basham:1978zq,Chen:2020vvp}
have emerged as a powerful tool to study the properties of strong
interactions, such as the precise extractions of the strong coupling
constant~\cite{CMS:2024mlf}.
\comment{From a theoretical viewpoint, these observables have inspired
  a thorough investigation of their field-theoretic
  properties~\cite{Hofman:2008ar,Korchemsky:2019nzm,Kologlu:2019mfz,Kologlu:2019bco,Chang:2020qpj,Chen:2020adz,Chen:2021gdk,Chang:2022ryc,Chen:2022jhb,Belitsky:2013ofa,Belitsky:2013bja,Belitsky:2013xxa,Dixon:2018qgp,Chen:2019bpb,Henn:2019gkr,Yan:2022cye,Lee:2022ige,Yang:2022tgm,Chicherin:2024ifn,Korchemsky:2021okt,Chicherin:2023gxt,Csaki:2024joe,Chen:2024iuv}.
  Owing to their simplicity, energy correlators inherit the quantum
  properties of the correlation functions, which encode fundamental
  information about the underlying field
  theory~\cite{Schwinger:1958qau}.
  This paves the way to new explorations of Quantum Chromodynamics
  using present and future collider data, as reflected in the wide
  phenomenological interest they have attracted in particle
  physics~\cite{Dixon:2019uzg,Kardos:2018kqj,Moult:2018jzp,Gao:2019ojf,Ebert:2020sfi,Gao:2020vyx,Komiske:2022enw,Holguin:2022epo,Neill:2022lqx,Chen:2022swd,Lee:2022ige,Ricci:2022htc,Craft:2022kdo,Duhr:2022yyp,Li:2021zcf,Jaarsma:2023ell,Lee:2023npz,Kang:2023big,Cao:2023qat,Gao:2023ivm,Xiao:2024rol,Chen:2023zlx},
  heavy-ion
  physics~\cite{Andres:2022ovj,Andres:2023xwr,Andres:2023ymw,Barata:2023zqg,Yang:2023dwc,Holguin:2023bjf,Barata:2023bhh}
  and nuclear
  physics~\cite{Liu:2022wop,Chen:2023wah,Cao:2023oef,Devereaux:2023vjz,Liu:2024kqt,Chen:2024nfl}.
  %
%  These aspects are reflected in the study of their theoretical
%  properties~\cite{Hofman:2008ar,Korchemsky:2019nzm,Kologlu:2019mfz,Kologlu:2019bco,Chang:2020qpj,Chen:2020adz,Chen:2021gdk,Chang:2022ryc,Chen:2022jhb,Belitsky:2013ofa,Belitsky:2013bja,Belitsky:2013xxa,Dixon:2018qgp,Chen:2019bpb,Henn:2019gkr,Yan:2022cye,Lee:2022ige,Yang:2022tgm,Chicherin:2024ifn,Korchemsky:2021okt,Chicherin:2023gxt,Csaki:2024joe,Chen:2024iuv},
%  which can be tested using present and future collider data hence
%  leading to a deeper understanding of Quantum Chromodynamics (QCD).
}
%
%More generally, they have attracted a wide phenomenological interest
%in particle
%physics~\cite{Dixon:2019uzg,Kardos:2018kqj,Moult:2018jzp,Gao:2019ojf,Ebert:2020sfi,Gao:2020vyx,Komiske:2022enw,Holguin:2022epo,Neill:2022lqx,Chen:2022swd,Lee:2022ige,Ricci:2022htc,Craft:2022kdo,Duhr:2022yyp,Li:2021zcf,Jaarsma:2023ell,Lee:2023npz,Kang:2023big,Cao:2023qat,Gao:2023ivm,Xiao:2024rol,Chen:2023zlx},
%heavy-ion
%physics~\cite{Andres:2022ovj,Andres:2023xwr,Andres:2023ymw,Barata:2023zqg,Yang:2023dwc,Holguin:2023bjf,Barata:2023bhh}
%and nuclear
%physics~\cite{Liu:2022wop,Chen:2023wah,Cao:2023oef,Devereaux:2023vjz,Liu:2024kqt,Chen:2024nfl}.
%
%\comment{REMOVE THIS -- These intertwine with the latest theoretical
%developments~\cite{Hofman:2008ar,Korchemsky:2019nzm,Kologlu:2019mfz,Kologlu:2019bco,Chang:2020qpj,Chen:2020adz,Chen:2021gdk,Chang:2022ryc,Chen:2022jhb,Belitsky:2013ofa,Belitsky:2013bja,Belitsky:2013xxa,Dixon:2018qgp,Chen:2019bpb,Henn:2019gkr,Yan:2022cye,Lee:2022ige,Yang:2022tgm,Chicherin:2024ifn,Korchemsky:2021okt,Chicherin:2023gxt,Csaki:2024joe,Chen:2024iuv},
%leading to a deeper understanding of field theories, notably of
%Quantum Chromodynamics (QCD).}

An $N$-point energy correlator is defined by weighing the cross section with the product of the energies of $N$ particles (e.g. within a jet), as a function of their relative angles.
One can define the corresponding projected $N$-point energy correlator ($\text{ENC}$)
by integrating the resulting correlator over these angles except for the largest one $\theta_L$,
as a univariate function of the angular variable $x_L\equiv (1-\cos\theta_L)/2$~\cite{Chen:2020vvp}.
In the collinear limit, considered in this Letter, one is interested in the regime in which such angular
variable is parametrically small, i.e. $\sqrt{x_L} Q\ll Q$, with $Q$ being the hard momentum transfer of the scattering process.
At leading power, the $\text{ENC}$ shows a classical scaling behavior
${\cal O}(1/x_L)$ in the collinear
limit~\cite{Hofman:2008ar,Dixon:2019uzg,Korchemsky:2019nzm,Kologlu:2019mfz,Chang:2020qpj,Chen:2020adz,Chen:2021gdk,Chang:2022ryc,Chen:2022jhb},
that is determined by Lorentz symmetry. This scaling is then violated
by quantum effects, which induce a mild additional dependence on the
hard scale $Q$. The evolution of $\text{ENC}$ with $Q$ can be
obtained using perturbative collinear resummation
techniques~\cite{Konishi:1979cb,Chen:2020vvp,Dixon:2019uzg,Karlberg:2021kwr,Lee:2022ige,vanBeekveld:2023lsa,Chen:2023zlx}.

The full exploitation of high-precision experimental data also demands
an understanding of the dynamics of $\text{ENC}$ beyond perturbation
theory.
A first-principle understanding of the deep non-perturbative limit in
which the angular distance $\sqrt{x_L}$ becomes of the order of the ratio
$\Lambda_{\rm QCD}/Q$, with $\Lambda_{\rm QCD}$ being a typical
hadronic scale, is currently out of reach.
Nevertheless, in the regime $1 \gg \sqrt{x_L}\gg \Lambda_{\rm QCD}/Q$,
one can approximate non-perturbative corrections in a power expansion
in $\Lambda_{\rm QCD}/Q$, gaining a better analytic control over their
properties. The next-to-leading power term of this expansion, commonly
denoted as \textit{power correction}, defines the leading
non-perturbative correction in this kinematic regime.
These power corrections can be studied using a range of analytic
techniques, which have been used to investigate observables belonging
to the $\text{ENC}$ family at lepton colliders both in the bulk of the
phase space~\cite{Korchemsky:1999kt,Belitsky:2001ij,Schindler:2023cww}
($x_L\neq 0,\,1$) as well as in the back-to-back
limit~\cite{Dokshitzer:1999sh} ($x_L\to 1$).

This Letter initiates a novel study of the $\text{ENC}$ in the
collinear ($x_L\to 0$) limit.
We will show that, similarly to their leading-power counterpart, the
coefficient of the linear ${\cal O}(\Lambda_{\rm QCD}/Q)$ power
correction to the $\text{ENC}$ exhibits a classical scaling behavior
fixed by symmetry arguments, that is violated at the quantum level in
a way that can be predicted using perturbation theory.
This phenomenon shares similarities with the violation of the well
known Bjorken
scaling~\cite{Bjorken:1968dy,Altarelli:1977zs,Gribov:1972ri,Dokshitzer:1977sg},
where the evolution of the non-perturbative structure functions with
the scale is fully
perturbative~\cite{Gribov:1972ri,Dokshitzer:1977sg,Altarelli:1977zs}.
Using the light-ray
OPE~\cite{Hofman:2008ar,Kologlu:2019mfz,Chang:2020qpj,Chen:2020adz,Chen:2021gdk,Chen:2023zzh},
we are able to calculate the evolution of the power correction with
the energy scale $Q$, hence predicting how they are related at
different scales.
This result marks a significant step forward in the theoretical
understanding of this class of observables beyond the perturbative
level.

\paragraph*{$\text{ENC}$  and the light-ray OPE.---}
The projected energy correlators ($\text{ENC}$) are correlation
functions of the energy flow operators ${\cal E}(n)$ in a physical
state
$|\Psi_q \rangle$~\cite{Basham:1978bw,Basham:1978zq,Hofman:2008ar},
defined as
$\langle {\cal E}(n_1) \cdots {\cal E}(n_k) \rangle_{\Psi_q} \equiv
\langle \Psi_q | {\cal E}(n_1) \cdots {\cal E}(n_k) | \Psi_q \rangle$,
where $q^\mu$ is the total momentum of the state $|\Psi_q \rangle$.
The energy flow operator ${\cal E}(n)$ is defined
as~\cite{Sveshnikov:1995vi,Hofman:2008ar}
\begin{equation}
  \label{eq:2}
  {\cal E}(n) = \mathbb{L}_{\tau=2}\left[ n^i  T_{0i}(t, r\vec{n})\right]  \,,
\end{equation}
where $T_{\mu\nu}$ is the energy-momentum tensor of QCD and the
operation $\mathbb{L}_{\tau}$ is the light
transform~\cite{Kravchuk:2018htv}
\begin{equation}
  \mathbb{L}_{\tau} = \lim_{r \to \infty} r^\tau \int_0^\infty dt \,.
\end{equation}
Examples for the state $|\Psi_q \rangle$ include those excited by the
electromagnetic current from the vacuum, the decay products of a Higgs
boson, or the scattering state of a high-energy collision.
Generic final states consist of the ensemble described by the density
$\rho_q=\sum_\Psi |\Psi_q \rangle \langle \Psi_q|$.

In recent years, the light-ray OPE has emerged as an efficient tool to
study energy correlators in the small angle limit. Originally
developed in the context of conformal collider
physics~\cite{Hofman:2008ar,Kologlu:2019mfz,Chang:2020qpj}, the
light-ray OPE has recently been found useful also in
QCD~\cite{Chen:2020adz,Chen:2021gdk,Chen:2023zzh}. For the simplest
two-point energy correlator~(EEC), the light-ray OPE at leading twist
reads
\begin{align}
  \label{eq:3}
\lim_{n_1 \to n_2}{\cal E}(n_1) {\cal E}(n_2) &= \frac{1}{x_L} \vec{C} \cdot \vec{\mathbb{O}}_{\tau = 2}^{[J=3]}(n_2)\\
& + \frac{\Lambda_{\rm QCD}}{x_L^{3/2}} \vec{D}\cdot \vec{\mathbb{O}}_{\tau = 2}^{[J=2]}(n_2) +\cdots \,, \notag
\end{align}
where $x_L = (n_1 \cdot n_2)/2$ is related to the angular distance of
two light-like directions $n_1$ and $n_2$, and $\vec{C}$,~$\vec{D}$
are dimensionless OPE coefficients.  The light-ray OPE formula in
\eqref{eq:3} describes the leading small-angle behavior of the
EEC. The operator $\vec{\mathbb{O}}_{\tau = 2}^{[J]}$ belongs to the
leading trajectory of the light-ray
operator~\cite{Kravchuk:2018htv}. For \textit{even} collinear spin
$J$, it can be obtained by a light transform of the following twist
$\tau=2$ local operators~\cite{Chen:2020adz,Chen:2021gdk}:
\begin{equation}
  \label{eq:4}
  \vec{\mathbb{O}}_{\tau=2}^{[J]}=\mathbb{L}_2[\vec{O}_{\tau = 2}^{[J]}]\,,\;\;
  \vec{O}_{\tau = 2}^{[J]} = \left(\! \begin{array}{c} \frac{1}{2^J} \bar \psi \gamma^+ (i D^+)^{J-1} \psi \\  \frac{-1}{2^J} G^{a, \mu +} (i D^+)^{J-2} G_{\mu}^{a,+} \end{array} \!\right) ,
\end{equation}
where $\gamma^+ = \bar{n} \cdot \gamma$, and $ \tau = \Delta - J$,
where $\Delta$ is the operator dimension. The energy flow operator
corresponds to the combination
${\cal E}=(1, 1) \cdot\vec{\mathbb{O}}_{\tau = 2}^{[J=2]}$, with
$J=2$.

\begin{figure}
  \centering
  \includegraphics[width=0.5\linewidth]{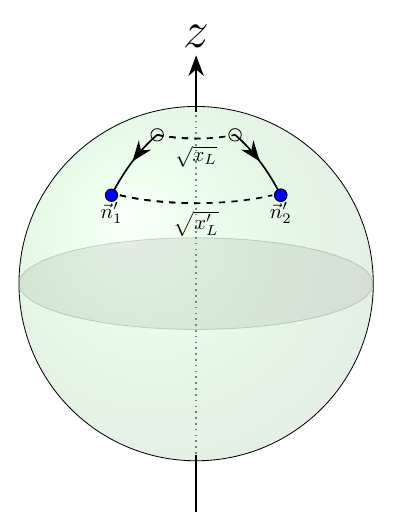}
  \caption{A boost in the positive $z$ direction increases the
    distance of two energy flow operator on the celestial
    sphere.}
  \label{fig:boost}
\end{figure}

The form of the light-ray OPE \eqref{eq:3} is determined by
dimensional analysis and Lorentz symmetry.
To understand this, we collect in TABLE~\ref{tab:1} the collinear spin
and dimension of all the ingredients entering the OPE~\eqref{eq:3}.
By dimensional analysis, imposing that the dimension of both sides
of~\eqref{eq:3} is the same fixes the dimension of the operators in
each term of the light-ray OPE. Imposing that the light-ray operators
have twist label $\tau=2$ fixes their collinear spin, which determines
the label $J=3$ for the first term in~\eqref{eq:3} and $J=2$ for the
${\cal O}(\Lambda_{\rm QCD}/Q)$ power correction.
This completely determines the classical scaling in $x_L$, e.g.
$x_L^{-3/2}$ for the ${\cal O}(\Lambda_{\rm QCD}/Q)$ term. The reason
is that the collinear spin of $x_L$ is completely fixed by its
transformation properties under a Lorentz boost, which acts as a
dilation of angles on the celestial sphere, as depicted in
FIG.~\ref{fig:boost}.
Using the quantities listed in TABLE~\ref{tab:1}, it is easy to verify
the legitimacy of \eqref{eq:3} (cf.~\cite{supplemental} for further
discussions).

\begin{table}
  \centering
  \begin{tabular}{c|c|c|c|c|c}
    \toprule
    & $\mathbb{L}_{\tau}$  & $\vec{O}_{\tau}^{[J]}$ & $\vec{\mathbb{O}}_{\tau}^{[J]}$ & $x_L$ & $\Lambda_{\rm QCD}$, $Q$ \\ 
    \midrule
    coll. spin & $1 - \tau $ & $-J$ & $1-(\tau + J) $ & $2$ & $0$\\
    dimension & $- \tau -1 $ & $\tau + J$ & $J-1$ & $0$ & $1$ \\
    \bottomrule
  \end{tabular}
  \caption{Collinear spin~(boost) and classical scaling dimension of
    various quantities appearing in the OPE~\eqref{eq:3}.}
  \label{tab:1}
\end{table}

The light-ray OPE in~\eqref{eq:3} can be generalized to higher point
correlators. For $N$ energy flow operators, the observable depends on
$N (N-1)/2$ angles for an isotropic state $| \Psi_q \rangle$. We are
interested in the projective $N$-point energy correlator, where the
higher dimensional distribution is projected to the axis of the
maximal angular distance of the $N$ energy flow operators. In this case
the OPE formula reads
\begin{multline}
  \label{eq:N_lightray_OPE}
  \int d \bar \Omega \lim_{n_i\to n} {\cal E}(n_1) \cdots {\cal E}(n_N) =  \frac{1}{x_L} \vec{C}_N \cdot \vec{\mathbb{O}}_{\tau = 2}^{[J=N+1]}(n) 
  \\
   + \frac{\Lambda_{\rm QCD}}{x_L^{3/2}} \vec{D}_N \cdot \vec{\mathbb{O}}_{\tau = 2}^{[J=N]}(n) + \cdots
\end{multline}
where all $n_i$ directions approach $n$. The angular integral is over
the $N (N-1)/2 - 1$ angles except the largest separation $x_L$ among
the $N$ directions. The general $N$-point projective energy correlator
can be defined as the normalized expectation value of the product of
$N$ energy flow operators in a state $|\Psi_q\rangle$:
\begin{equation}
  \text{ENC}_{\Psi_q}(x_L, Q) = \frac{4\pi}{\sigma_{\Psi_q} Q^N} \int d \bar \Omega \lim_{n_i\to n}  \langle  {\cal E}(n_1) \cdots {\cal E}(n_N) \rangle_{\Psi_q} \,,
\end{equation}
where $\sigma_{\Psi_q} = \langle \Psi_q | \Psi_q \rangle$ and $Q=q^0$
is the energy of the state, e.g. the center-of-mass energy in
$\gamma^{*}\to q\bar{q}$ or $h\to gg$, or the energy of jets.

For $N=2$ it reduces to the conventional
EEC~\cite{Basham:1978bw,Basham:1978zq}.
In parton language, the first term in the r.h.s. of Eq.~\eqref{eq:3}
gives rise to the factorisation theorem of
Refs.~\cite{Dixon:2019uzg,Chen:2020vvp}, known at
Next-to-Next-to-Leading Logarithmic (NNLL) accuracy in
QCD~\cite{Dixon:2019uzg,Chen:2023zlx}.
The operators $\vec{\mathbb{O}}_{\tau = 2}^{[J=3]}$ are mapped onto
the hard function while the OPE coefficients are encoded in the jet
function.
The leading-power EEC exhibits a classical scaling behavior
${\cal O}(1/x_L)$, which is mildly violated by quantum corrections
that modify its dependence on the energy $Q$.
Analogous considerations hold for the $\text{ENC}$ starting from
Eq.~\eqref{eq:N_lightray_OPE}.

We now focus on the power correction to the $N$-point projective
energy correlator by defining
\begin{equation}
  \label{eq:7}
  \text{ENC}_{\Psi_q}^{\rm N.P.}(x_L, Q) \equiv \text{ENC}_{\Psi_q} (x_L, Q) -  \text{ENC}_{\Psi_q}^{\rm P.T.}(x_L, Q)\,,
\end{equation}
where the subscript P.T. denotes the leading power, perturbative part
of the energy correlator.
Eq.~\eqref{eq:N_lightray_OPE} predicts a classical scaling behavior
${\cal O}(x_L^{-3/2})$ for the leading power correction.
The assumption of linearity in $\Lambda_{\rm QCD}$ is supported by
predictions from hadronization~\cite{Basham:1978zq,Dokshitzer:1995qm},
Wilson loop~\cite{Korchemsky:1999kt,Belitsky:2001ij} and
renormalon~\cite{Schindler:2023cww} models.

\begin{figure}
  \includegraphics[width=0.9\linewidth]{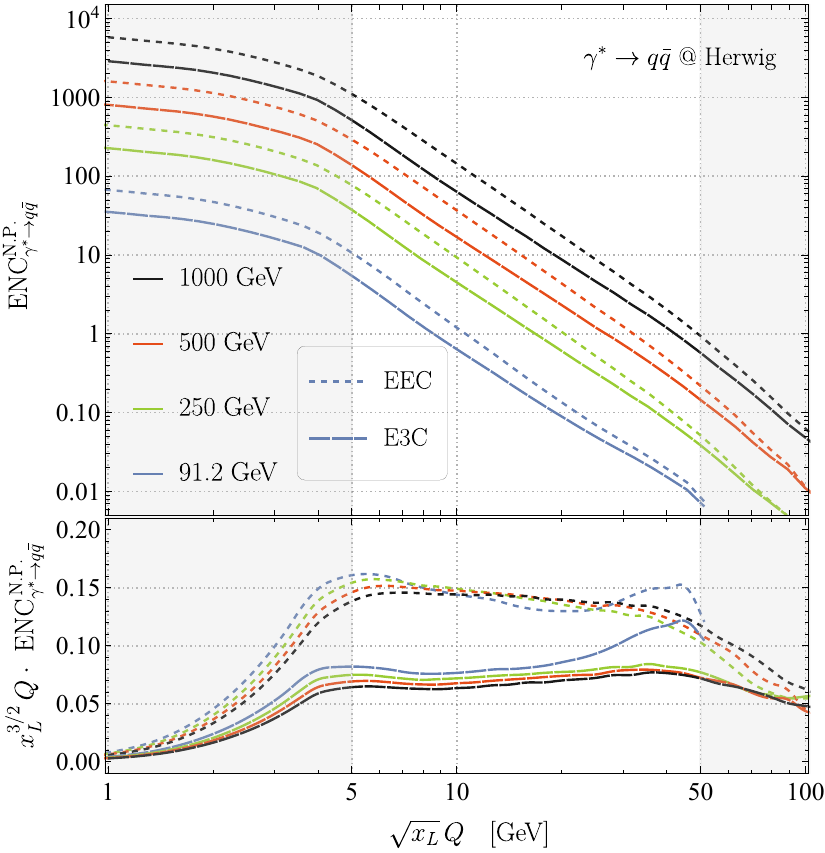}
  \caption{The upper panel shows the classical scaling of the energy
    correlators across energy and angular scales. The lower panel
    highlights the mild quantum scaling violation.}
  \label{fig:scaling}
\end{figure}

The classical scaling can be verified using \comment{Monte Carlo
  simulations} for $e^+ e^-$ collision at different $Q$, as shown in
FIG.\ref{fig:scaling} for the process $\gamma^*\to q\bar{q}$.
In the following we consider the regime
$Q\gg Q \sqrt{x_L}\gg \Lambda_{\rm QCD}$, shown in the unshaded region
of the plot, where the scaling is clearly visible.

\paragraph*{Scaling violation in power corrections.---}

The above discussion about classical scaling is based only on Lorentz
symmetry and classical dimensional analysis.
We now show that the OPE also predicts quantum corrections that
slightly violate this scaling in the perturbative region
$Q\gg Q \sqrt{x_L}\gg \Lambda_{\rm QCD}$.
We start by factoring out the classical scaling of
$\text{ENC}_{\Psi_q}^{\rm N.P.}(x_L, Q)$, and write
\begin{equation}\label{eq:ENC-NP}
  \text{ENC}_{\Psi_q}^{\rm N.P.}(x_L, Q) = \frac{  \text{ENC}_{1,\Psi_q}^{\rm
      N.P.}(K_\perp, Q)}{x_L^{3/2}Q}+\dots\,,
\end{equation}
where $K_\perp = \sqrt{x_L} Q$ characterizes the exchanged transverse
momentum scale and we neglected subleading power corrections in
the r.h.s.
Classically, no dependence of
$ \text{ENC}^{\rm N.P.}_{1,\Psi_q} (K_\perp, Q)$ on $Q$ is
expected. At the quantum level, the function
$ \text{ENC}^{\rm N.P.}_{1,\Psi_q} (K_\perp, Q)$ mildly depends on
$Q$.
This can be appreciated in the lower panel of FIG.~\ref{fig:scaling},
where the classical scaling has been removed.
We refer to the small residual dependence of
$ \text{ENC}^{\rm N.P.}_{1,\Psi_q} (K_\perp, Q)$ on $Q$ as to quantum
{\it scaling violation}.% phenomenon.

We now show how the scaling violation is caused by the renormalization
group (RG) evolution of $\vec{\mathbb{O}}_{\tau = 2}^{[J]}$. The
light-ray OPE \eqref{eq:N_lightray_OPE} provides the following
factorized prediction for $\text{ENC}_{1,\Psi_q}^{\rm N.P.}$
\begin{multline}
  \label{eq:EEC_power_corr}
  \text{ENC}_{1,\Psi_q}^{\rm N.P.}(K_\perp, Q) = \Lambda_{\rm QCD}
  \\
    \times \vec{D}_N \left(\frac{K_\perp^2}{\mu^2}, \frac{\Lambda_{\rm QCD}^2}{\mu^2} \right) \cdot 
   \frac{\langle  \vec{\mathbb{O}}_{\tau = 2}^{[J=N]}(n;\mu) \rangle_{\Psi_q}}{{(4\pi)^{-1}\sigma_{\Psi_q}   Q^{N-1}}} \left( \frac{Q^2}{\mu^2}\right)\,,
\end{multline}
where $\mu$ is a factorization scale that separates the perturbative
matrix element of light-ray operators and the non-perturbative OPE
coefficients $\vec{D}_N$, which also depend on $\Lambda_{\rm QCD}$.
The light-ray operator $\vec{\mathbb{O}}_{\tau = 2}^{[J]}(n;\mu)$
satisfies the DGLAP
equation~\cite{Chen:2020adz,Chen:2021gdk,Chen:2023zzh,Caron-Huot:2022eqs}
\begin{equation}\label{eq:light-ray_RG}
\mu \frac{d}{d\mu}  \vec{\mathbb{O}}_{\tau = 2}^{[J]}(n;\mu)= \gamma_{\tau=2}^{[J]}(\mu)\cdot  \vec{\mathbb{O}}_{\tau = 2}^{[J]}(n;\mu)\,,
\end{equation}
where $\gamma_{\tau = 2}^{[J=N]}$ is the anomalous dimension matrix of
the twist-2 light-ray operators, which admits the perturbative expansion
$ \gamma_{\tau = 2}^{[J]}(\mu) = \sum_{k=0}^\infty
\left(\frac{\alpha_s(\mu)}{4 \pi} \right)^{k+1} \gamma_{\tau =
  2}^{[J],(k)}$.
The leading order expression $\gamma_{\tau = 2}^{[J],(0)}$ can be
found in~\cite{Gross:1974cs,Gross:1973ju}, while higher-order
calculations of $\gamma_{\tau = 2}^{[J]}$ and their inverse Mellin
transform can be found
in~\cite{Moch:2004pa,Vogt:2004mw,Chen:2020uvt,Blumlein:2021enk,Gehrmann:2023iah,Moch:2021qrk,Falcioni:2023luc,Falcioni:2023vqq,Falcioni:2024xyt}.
Renormalization group invariance implies that
\begin{equation}
  \label{eq:RG_invariance}
  \mu \frac{d \vec{D}_N}{d \mu}
  = - \vec{D}_N \cdot \gamma_{\tau = 2}^{[J=N]} \,.
\end{equation}
By observing that $\vec{D}_N$ does not depend on $Q$ explicitly,
Eq.~\eqref{eq:EEC_power_corr} suggests that one can predict the $Q$
dependence of $\text{ENC}_{1,\Psi_q}^{\rm N.P.}$ at \textit{fixed}
$K_\perp$ solely from the $Q$ dependence of the matrix element of
light-ray operators.
Specifically, we let $\mu = K_\perp$ in the OPE coefficient, and
evolve the matrix element of the light-ray operators from $Q$ to
$K_\perp$ using the renormalization group equation
\eqref{eq:light-ray_RG}. Since the physical state $|\Psi_q\rangle$ is
$\mu$ independent, at fixed $K_\perp$ the $Q$ dependence of
$\text{EEC}_{1,\Psi_q}^{\rm N.P.}$ is given by
\begin{align}
  \label{eq:evolution}
&  \text{ENC}_{1,\Psi_q}^{\rm N.P.}(K_\perp, Q) = \Lambda_{\rm QCD}
  \\
  &\times \vec{D}_N \left(1, \frac{\Lambda_{\rm QCD}^2}{K_\perp^2} \right) \cdot U_N(K_\perp, Q)    \cdot  
  \frac{\langle  \vec{\mathbb{O}}_{\tau = 2}^{[J=N]}(n;Q) \rangle_{\Psi_q} }{(4\pi)^{-1}\sigma_{\Psi_q} Q^{N-1}} \,,\notag
\end{align}
where $U_N(K_\perp, Q)$ is the evolution operator
\begin{equation}\label{eq:RG_kernel}
  U_N(K_\perp, Q) \equiv \mathbb{P} \exp \left( - \int_{K_\perp}^Q
    \frac{d \mu}{\mu} \gamma_{\tau = 2}^{[J=N]}(\mu ) \right) \,.
\end{equation}
This is one of the main results of this Letter.

\begin{figure}[htbp]
\begin{center}
\includegraphics[width=0.9\linewidth]{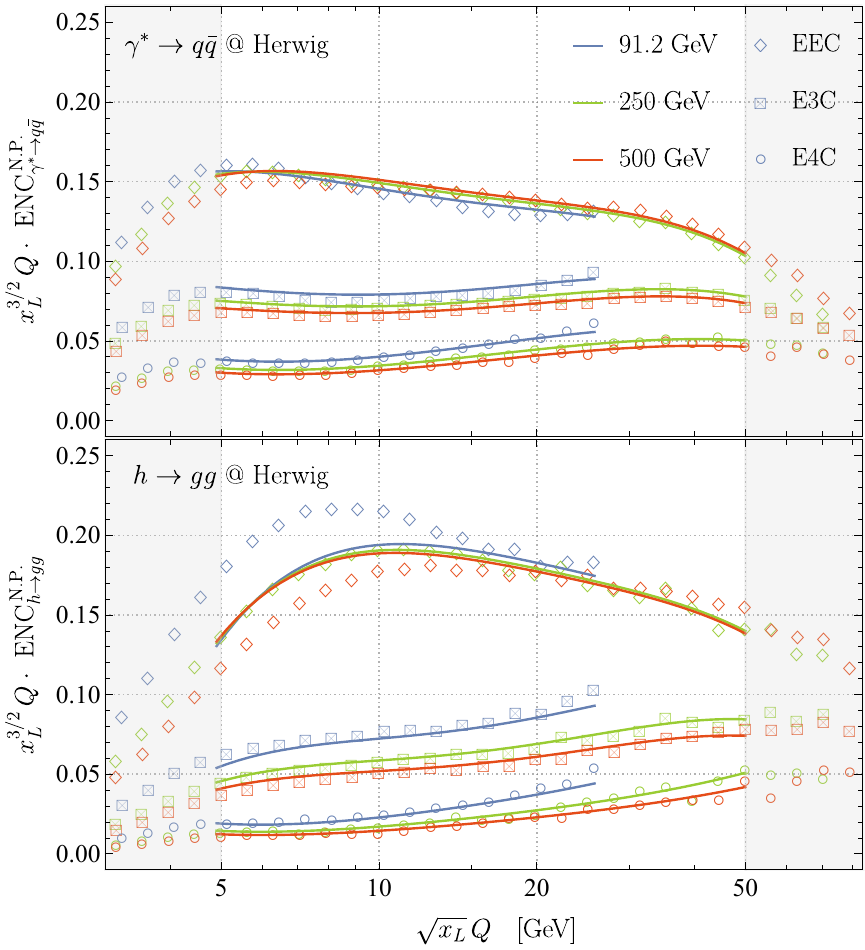}
\caption{Comparison of analytic and Monte Carlo predictions for the
  quantum scaling violation for $N$-point projected correlators in
  $\gamma^*\to q\bar{q}$ and $h\to gg$.}
\label{fig:ee_plot}
\end{center}
\end{figure}

We finally discuss the implications of Eq.~\eqref{eq:evolution} for
the factorisation formula of the projected
correlator~\cite{Dixon:2019uzg,Chen:2020vvp}:
\begin{equation}
\label{eq:factorization_formula}
\text{ENC}_{\Psi_q}(x_L, Q) = \int_0^1 \!\!\!dx  \frac{x^N}{x_L} \vec{J}_N(\frac{x_L x^2  Q^2}{\mu^2}, \frac{\Lambda_{\rm QCD}^2}{\mu^2}) \cdot \vec{H}(x,\frac{Q}{\mu})\,.
\end{equation}
Based on the equivalence~\cite{Chen:2023zzh} of the light-ray OPE and
Eq.~\eqref{eq:factorization_formula}, we can extend the above
factorization formula to include the power corrections derived in this
Letter.
The hard function $\vec{H}$ encodes the probability of producing a
parton with energy fraction $x$ and it corresponds to the normalized
matrix element of the twist $\tau=2$ operators in
Eq.~\eqref{eq:N_lightray_OPE}. The jet function $\vec{J}_N$ is
sensitive to the fragmentation at small angular scales and, as such,
it encodes also the non-perturbative dynamics. It is
mapped~\cite{Chen:2023zzh} onto the OPE coefficients of
Eq.~\eqref{eq:N_lightray_OPE}, from which we can deduce the following
expansion
\begin{multline}
\vec{J}_N (\frac{x_L x^2  Q^2}{\mu^2}, \frac{\Lambda_{\rm QCD}^2}{\mu^2}) 
= \vec{J}_N^{\rm P.T.}(\frac{x_L x^2  Q^2}{\mu^2}, \alpha_s(\mu)) \\
+ \frac{\Lambda_{\rm QCD}}{x \sqrt{x_L} Q} \vec{J}^{(1)}_N(\frac{x_L x^2  Q^2}{\mu^2}, \frac{\Lambda_{\rm QCD}^2}{\mu^2})+\cdots\,,
\end{multline}
where $\vec{J}^{\rm P.T.}_N$ is the perturbative jet
function~\cite{Chen:2020vvp,Dixon:2019uzg,Lee:2022ige,Chen:2023zlx}
and $\vec{J}^{(1)}_N = \vec{D}_N \big|_{K_\perp \to x K_\perp}$ is the
corresponding power correction.

\begin{figure}[htbp!]
\begin{center}
  \includegraphics[width=0.9\linewidth]{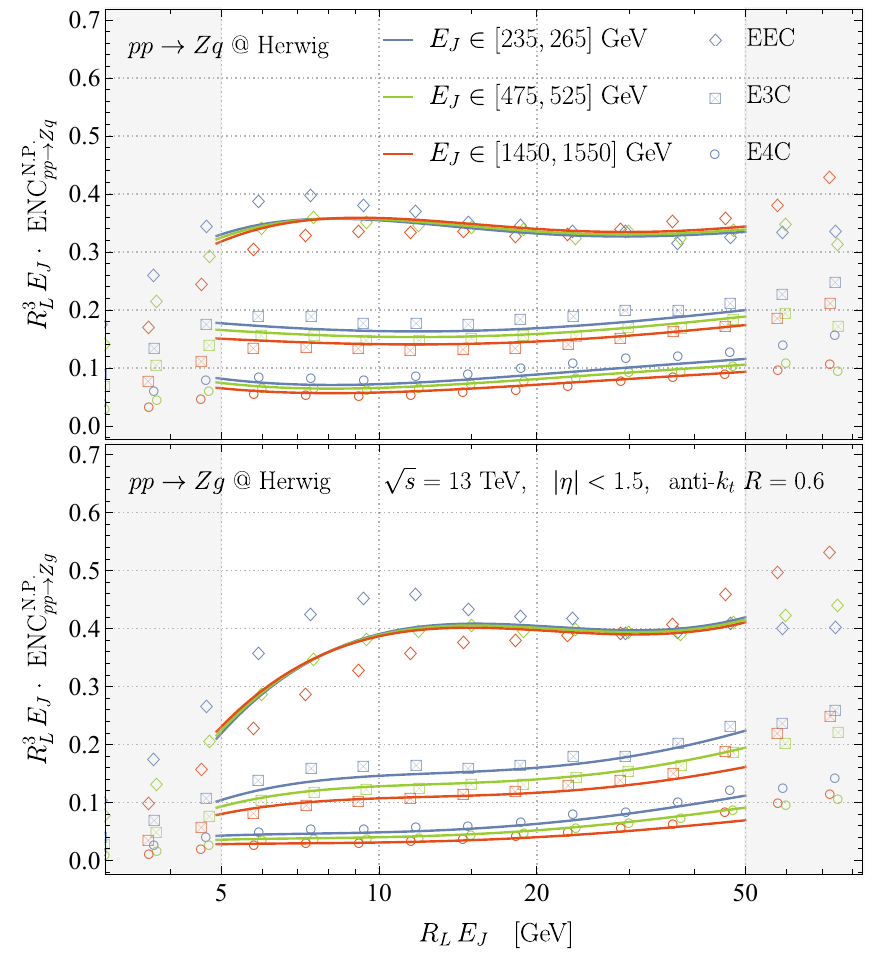}
  \caption{Comparison of analytic and Monte Carlo predictions for the
    quantum scaling violation for $N$-point projected correlators in
    $pp\to Z q$ and $p p \to Z g$.}
\label{fig:pp_plot}
\end{center}
\end{figure}

\paragraph*{Monte Carlo validation.---}\label{sec:MC}
We can now explicitly use Eq.~\eqref{eq:evolution} to relate the power
correction at a reference scale $Q_0$ to that at a scale $Q$. We can
express the solution in terms of two non-perturbative functions of
$K_\perp$ defining the two components of $\vec{D}$ at a reference
scale $Q_0$, which can be extracted from the fragmentation of quarks
and gluons. At the leading-logarithmic order we find
(cf.~\cite{supplemental} for details)
\begin{equation}
\label{eq:LP_ee_higgs}
\begin{pmatrix}
  \text{ENC}_{1,\gamma^*\to q\bar{q}}^{\rm N.P.}(Q)\\
    \text{ENC}_{1,h\to gg }^{\rm N.P.}(Q)
\end{pmatrix}^T\!\!\!
=
\begin{pmatrix}
  \text{ENC}_{1, \gamma^*\to q\bar{q}}^{\rm N.P.}(Q_0)\\
    \text{ENC}_{1, h\to gg}^{\rm N.P.}(Q_0)
\end{pmatrix}^T
\!\!\! \cdot\, U^{\rm LL}_N(Q_0, Q)\,,
\end{equation}
where $K_\perp$ is fixed and kept implicit and
$U_N^{\rm LL}(Q_0,Q) =
\left[\alpha_s(Q)/\alpha_s(Q_0)\right]^{\gamma_{\tau=2}^{[N],(0)}/(2\beta_0)
}$.

We extract the functions
$\text{ENC}_{1, \gamma^*\to q\bar{q}}^{\rm N.P.}(Q_0)$ and
$\text{ENC}_{1, h\to gg}^{\rm N.P.}(Q_0)$ from $\gamma^*\to q\bar{q}$
and $h\to gg$ at $Q_0=250$\GeV for 2-, 3- and 4-point correlators and
predict their distribution at a different c.o.m. energy
$Q\in 91.2-500$\GeV.
Specifically, we use events generated with
MadGraph5~\cite{Alwall:2011uj}, showered with Herwig
7.2~\cite{Bellm:2019zci}~\footnote{\comment{Specifically, we use the
  dot-product preserving shower and corresponding tune from
  Ref.~\cite{Bewick:2019rbu}.}} and analyzed with
Rivet~\cite{Bierlich:2019rhm}.\footnote{We have repeated the analysis
  also with Herwig 7.3~\cite{Bewick:2023tfi} and
  Pythia8~\cite{Sjostrand:2014zea}, finding consistent results.}
The results are shown in FIG.~\ref{fig:ee_plot}, which displays a
comparison of Eq.~\eqref{eq:LP_ee_higgs} to the \comment{Monte Carlo
  prediction} obtained with Eqs.~\eqref{eq:7}
and~\eqref{eq:ENC-NP}. We notice that the latter contains subleading
power corrections not accounted for in Eq.~\eqref{eq:LP_ee_higgs}.
In general, we observe very good agreement, hence validating the
expectation for the perturbative scaling violation presented in this
Letter.
From FIG.~\ref{fig:ee_plot} we observe that in the case of the EEC the
region of validity of Eq.~\eqref{eq:LP_ee_higgs} is substantially
pushed towards larger angles.
An explanation of this fact, particularly prominent in the gluonic
case, is that subleading power corrections neglected in the
OPE~\eqref{eq:3} receive a contribution from operators with
$J \sim 1$,
whose anomalous dimensions feature a strong enhancement due to the
radiation of soft
gluons~\cite{Lipatov:1996ts,Brower:2006ea,Jaroszewicz:1982gr,Kotikov:2000pm,Kotikov:2002ab}. The
enhanced quadratic power corrections may be ultimately responsible for
the discrepancy in the left region of the plot. This phenomenon is
present only in the EEC case while for $N>2$ the contribution of $J<2$
operators is further power suppressed.

It is interesting to apply the same procedure to the case of
$\text{ENC}$ measured on hadronic jets at the LHC~\cite{CMS:2024mlf},
in the limit in which the largest angular resolution
$\sqrt{x_L}\to R_L\equiv \sqrt{\Delta y^2+\Delta \phi^2}$ between
detectors is much smaller than the jet radius $R$. We consider
$ p p\to Z+q/g$ jets LHC events at $\sqrt{s}=13$ TeV, with jet
energies in the range $E_J\in 250-1500$ GeV. Accordingly we now define
$K_\perp=R_L E_J$.
Jets are defined using the anti-$k_t$ algorithm~\cite{Cacciari:2008gp}
with a jet radius $R=0.6$, as implemented in
FastJet~\cite{Cacciari:2011ma}.
\comment{We generate separately events with quark and gluon jets, that
  we extract from $Z q$ and $Z g$ final states, respectively. An
  analysis at higher perturbative orders, however, would require a
  more refined definition of quark and gluon jet fractions.}
The results are shown in FIG.~\ref{fig:pp_plot}, where the solid lines
indicate our predictions from Eq.~\eqref{eq:LP_ee_higgs} with
$Q_0=500$\GeV and $Q=E_J$.
The effect of initial-state radiation, present in $pp$ collisions,
impacts mildy the correlators measured inside jets at the
non-perturbative level (e.g. via colour reconnection).
While the EEC, as in FIG.~\ref{fig:ee_plot}, is affected by large
subleading power corrections, the analytic prediction describes very
well the \comment{simulation} for the $N>2$ correlators, confirming
the validity of our results also in the hadron-collider case
(cf.~\cite{supplemental} for additional studies).

We envision that the leading power correction can be directly
extracted from experimental data at a reference scale and then evolved
at different scales using the results presented in this letter.
In Ref.~\cite{supplemental} we present also a study of the effect of
quantum scaling violation on ratios of energy correlators, used to
measure $\alpha_s$ in Ref.~\cite{CMS:2024mlf}.
This work will enhance the role of energy correlators in the precision
physics programme and their use for the extraction of fundamental
properties of QCD at the LHC and future colliders.

\paragraph*{Note added.---}
While this article was being completed, Ref.~\cite{Lee:2024esz}
presented a related study of the power corrections to the projected
correlators using a renormalon analysis in the context of extractions of
the strong coupling constant.
The connection of their findings to our prediction from the light-ray
OPE is non-trivial and deserves further investigation.

\begin{acknowledgments}
  We thank Yibei Li, Kyle Lee, Aditya Pathak, David Simmons-Duffin,
  Iain Stewart, Zhiquan Sun, Gherardo Vita, Xin-Nian Wang and
  Alexander Zhiboedov for discussions and Silvia Ferrario Ravasio for
  assistance with the Herwig7 event generator. HC wishes to thank the
  theory group at CERN for hospitality where this work was
  initiated. PM wishes to thank the Center for High Energy Physics of
  Peking University for hospitality while part of this work was
  carried out.
  The work of HC, ZX, and HXZ was supported by the National Natural
  Science Foundation of China (NSFC) under Grant No. 11975200 and No. 12425505. HXZ is
  also supported by the Startup Grant from Peking University and the
  Asian Young Scientist Fellowship.
  The work of PM is funded by the European Union (ERC, grant agreement
  No. 101044599).
  Views and opinions expressed are however those of the authors only
  and do not necessarily reflect those of the European Union or the
  European Research Council Executive Agency. Neither the European
  Union nor the granting authority can be held responsible for them.
\end{acknowledgments}

\bibliographystyle{apsrev4-2}
\bibliography{refs}

\input{supplementary_material}

\end{document}

%% file: supplementary_material.tex
\newpage

\onecolumngrid
\newpage
\appendix

\makeatletter
\renewcommand\@biblabel[1]{[#1S]}

\renewcommand{\theequation}{S.\arabic{equation}}
\setcounter{equation}{0}
\makeatother

%======================================================================
\section*{Supplemental material}

%%%%%%%%%%%%%%%%%%%%%%%%%%%%%%%%%%%%%%%%%%%%%%%%%%%%%%%
\subsection{$\text{ENC}$ measured in jets in $\gamma^*\to q\bar{q}$ and $h\to gg$}
\label{app:enc-jets}
In this appendix we repeat the analysis shown in the main text for the case of $\text{ENC}$  measured inside jets produced in the decays $\gamma^*\to q\bar{q}$ and $h\to gg$.
For consistency with the $pp\to Z+$jets case shown in
FIG.~\ref{fig:pp_plot}, we consider the $e^+e^-$ version of the
anti-$k_t$ algorithm, belonging to the family of generalized $k_t$
algorithms in Ref.~\cite{Cacciari:2011ma}, with $R=0.6$, and denote
the largest angular distance by $R_L\ll R$.
\begin{figure}[h!]
\begin{center}
\includegraphics[width=0.5\linewidth]{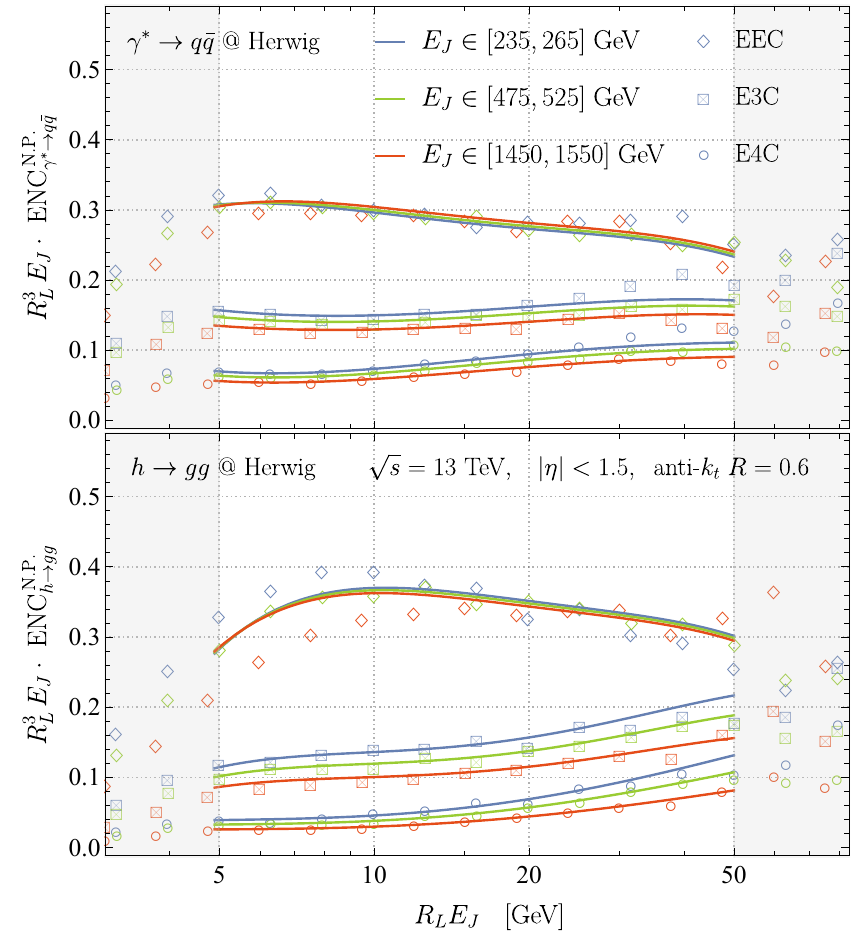}
\caption{Comparison of analytic and Monte Carlo predictions for the
  quantum scaling violation for $N$-point projected correlators measured in jets in
  $\gamma^*\to q\bar{q}$ and $h\to gg$.}
\label{fig:eeJet_plot}
\end{center}
\end{figure}

The non-perturbative boundary conditions are extracted at
$Q_0=500$\GeV and then Eq.~\eqref{eq:LP_ee_higgs} is used to predict
the power corrections in the range $E_J\in 250-1500$ GeV. The results
are summarised in FIG.~\ref{fig:eeJet_plot}.
When compared to FIG.~\ref{fig:pp_plot}, we observe a better agreement between the analytic prediction and the Monte Carlo data for the fermionic final state. This is due to the absence of initial-state radiation, which in FIG.~\ref{fig:pp_plot} introduces an additional source of non-perturbative correlation with the radiation within the jet.
The description of Monte Carlo data is somewhat less good in the right
region of the plot, where effects due to the finite size of the jet
play a role.

\subsection{Properties of Light-ray Operators and ansatz for the Light-ray OPE}

In this appendix we elaborate on the {\it classical} symmetry properties of light-ray operators and their guidance to build a general light-ray OPE ansatz for non-perturbative power corrections. We discuss the case of $d=4$ spacetime dimensions, but the generalization to other spacetime dimensions is straightforward.

Let us consider a light-ray operator $\mathbb{O}_\tau^{[J]}$ as the light transform of the corresponding local operator $\mathcal{O}_\tau^{[J]}(x;\bar{n})=\mathcal{O}_\tau^{\mu_1\cdots \mu_J}(x) \bar{n}_{\mu_1} \cdots \bar{n}_{\mu_J}$
\begin{equation}
\mathbb{O}^{[J]}_\tau (n) = \mathbb{L}_\tau [\mathcal{O}_\tau^{[J]}(x;\bar{n})\big|_{x=(t,r\vec{n})}]= \lim_{r\to \infty} r^\tau \int_0^\infty dt \, \mathcal{O}_\tau^{[J]}(x;\bar{n})\big|_{x=(t,r\vec{n})} \,,
\end{equation}
where we define two null vectors $n^\mu=(1,\vec{n})$, $\bar{n}^\mu =(1,-\vec{n})$. The local operator $\mathcal{O}_\tau^{[J]}(x;\bar{n})$ has dimension $\Delta=\tau+J$ and the light transform operator $\mathbb{L}_\tau$ carries dimension $-(1+\tau)$, which leads to the fact that the light-ray operator $\mathbb{O}^{[J]}_\tau (n)$ has dimension $J-1$.
The collinear spin is associated with Lorentz boosts along $\vec{n}$. To count the collinear spin, we choose the convention to assign $n^\mu$ with collinear spin $+1$, and $\bar{n}^\mu$ with $-1$. To associate a collinear spin to the light transform operation, we rewrite it in terms of light-cone coordinates
\begin{equation}
\mathbb{L}_\tau = \lim_{\bar{n}\cdot x \to \infty} \left(\frac{\bar{n}\cdot x}{2}\right)^\tau \int_{-\infty}^{\infty} d(n\cdot x)\,,
\end{equation}
from which we can directly identify the collinear spin to be $1-\tau$. Since the local operator $\mathcal{O}_\tau^{[J]}(x;\bar{n})$ carries collinear spin $-J$, the collinear spin of the light-ray operator $\mathbb{O}^{[J]}_\tau (n)$ is $1-(\tau+J)$.

One reason to discuss dimension and collinear spin is that the expansion parameters $\zeta_{ij}=\frac{n_i\cdot n_j}{2}$ and $x_L=\max{\zeta_{ij}}$ in the small angle limit are dimensionless and have nice transformation properties under boost. Assume we make an infinitesimal boost transformation along $\vec{n}$. The unit vectors $\vec{n}_i$ on the celestial sphere change in the following way
\begin{equation}
\vec{n}_i \to \vec{n}_i + \epsilon (\vec{n}-(\vec{n}\cdot \vec{n}_i)\vec{n}_i)\,,
\end{equation}
where $\epsilon$ is the infinitesimal parameter of the boost transformation. This leads to the transformation of $\zeta_{ij}$
\begin{equation}
\zeta_{ij}\to (1-\epsilon(\vec{n}\cdot \vec{n}_i + \vec{n}\cdot \vec{n}_j))\zeta_{ij} \approx (1-2\epsilon)\zeta_{ij}\,,
\end{equation}
where we have used the collinear approximation $\vec{n}\cdot\vec{n}_i\approx \vec{n}\cdot\vec{n}_j\approx 1$ in the second step. This means that the Lorentz boost acts as a dilation on the celestial sphere. As suggested by the coefficient of $\epsilon$, in the collinear limit, $\zeta_{ij}$ and $x_L$ carry weight $2$ under a boost transformation.

Now let us consider the collinear limit $n_i\to n$ for the product of $N$ energy flow operators in terms of light-ray OPE
\begin{equation}
\label{eq:light-ray_ansatz}
\lim_{n_i\to n}    {\cal E}(n_1)\cdots {\cal E}(n_N) \supset  \Lambda_{\rm QCD}^\alpha \mathcal{C}_{\alpha, \tau, J}(\{\zeta_{ij}\}) \mathbb{O}_{\tau}^{[J]}(n)\,,
\end{equation}
where we consider the OPE coefficient with dimension $\alpha$, with $\mathcal{C}_{\alpha, \tau, J}$ being a dimensionless function. The l.h.s. of \eqref{eq:light-ray_ansatz} has dimension $N$, while the r.h.s. has the dimension $\alpha+(J-1)$.
Balancing the dimension on both sides  fixes the label $J$ to be
\begin{equation}
J=N+1-\alpha\,.
\end{equation}

Note that the energy flow operator $\mathcal{E}$ has the labels $\tau=2$, $J=2$ and hence has collinear spin $-3$. In the collinear limit, the collinear spin of \eqref{eq:light-ray_ansatz} on the l.h.s. is then $-3N$ and the collinear spin of $\mathbb{O}_{\tau}^{[J=N+1-\alpha]}(n)$ is $\alpha-\tau-N$. This requires that the coefficient function $\mathcal{C}_{\alpha, \tau, J=N+1-\alpha}(\{\zeta_{ij}\})$ has collinear spin $\tau-2N-\alpha$, that is
\begin{equation}
\mathcal{C}_{\alpha, \tau, J=N+1-\alpha}(\{\zeta_{ij}\})=\frac{1}{x_L^{(\alpha+2N-\tau)/2}} \widetilde{\mathcal{C}}_{\alpha, \tau, J=N+1-\alpha}(\{\zeta_{ij}/x_L\})\,,
\end{equation}
where we have factored out the overall $x_L$ dependence and $\widetilde{\mathcal{C}}_{\alpha, \tau, J=N+1-\alpha}(\{\zeta_{ij}/x_L\})$, carrying the information of the shape dependence on the $N$-point configuration, then has collinear spin $0$. For projected $N$-point energy correlators, additional integrations over the angular variables will introduce $x_L$ dependence due to the boundary constraint $x_L=\max \zeta_{ij}$. Without loss of generality, let us assume $x_L = \zeta_{12}$ and evaluate the remaining $N-2$ angular integrals with the measure $d^2\vec{n}_3\cdots d^2\vec{n}_N$. Working out the explicit Jacobian from the change of measure is not easy, but the overall scaling with respect to $x_L$ is straightforward
\begin{equation}
\int_{\zeta_{ij}\leq x_L} d^2\vec{n}_3\cdots d^2\vec{n}_N \;\sim\; x_L^{N-2}\,.
\end{equation}
Therefore, we obtain the following ansatz for a term of dimension $\alpha$ contributing to the light-ray OPE 
\begin{equation}
\label{eq:light-ray_OPE_tau}
\int d\Omega \lim_{n_i\to n}    {\cal E}(n_1)\cdots {\cal E}(n_N) \supset \Lambda_{\rm QCD}^\alpha \frac{D_{N,\tau}^{(\alpha)}}{x_L^{(\alpha-\tau)/2+2}} \mathbb{O}_{\tau}^{[J=N+1-\alpha]}(n)\,,
\end{equation}
where the coefficient $D_{N,\tau}^{(\alpha)}$ is related to $\widetilde{\mathcal{C}}_{\alpha, \tau, J=N+1-\alpha}(\{\zeta_{ij}/x_L\})$ by integrating out all the shape dependence.

Eq. \eqref{eq:light-ray_OPE_tau} shows that the expansion in the collinear limit is controlled by the twist of light-ray operators. In perturbative QCD, the leading twist is $2$ in which the classical degeneracy of quark and gluon twist-2 operators occurs. We can now neglect, in good approximation, the contribution from higher twist operators to the collinear limit and we obtain the following form for the leading twist light-ray OPE
\begin{align}
\label{eq:light-ray_OPE_int}
\int d\Omega \lim_{n_i\to n}    {\cal E}(n_1)\cdots {\cal E}(n_N) 
= \int_{0}^{N-\epsilon} d\alpha \frac{\Lambda_{\rm
  QCD}^\alpha}{x_L^{1+\alpha/2}} \vec{D}_{N,\tau=2}^{(\alpha)} \cdot
  \vec{\mathbb{O}}_{\tau = 2}^{[J=N+1-\alpha]}(n) + o(\Lambda_{\rm QCD}^{N-\epsilon}) + \text{higher twists}\,,
\end{align}
where the integral over the dimension parameter $\alpha$ indicates that all twist-2 light-ray operators can in principle appear in the light-ray OPE. We introduce a small positive parameter $\epsilon>0$ to avoid the complication of non-DGLAP evolution when the label $J\sim 1$~\cite{Lipatov:1996ts,Brower:2006ea,Jaroszewicz:1982gr,Kotikov:2000pm,Kotikov:2002ab}.
The perturbative result corresponds to the case where
$\vec{D}^{(\alpha)}_{N,\tau=2}$ behaves like $\delta(\alpha)$ and its
derivatives near $\alpha\sim 0$~\cite{Chen:2023zzh}.  In
Eqs.~\eqref{eq:3} and \eqref{eq:N_lightray_OPE}, we supplement
Eq.~\eqref{eq:light-ray_OPE_int} with the assumption that the leading
power correction is of order $\mathcal{O}(\Lambda_{\rm QCD})$, which
translates to the condition that $\vec{D}^{(\alpha)}_{N,\tau=2}$ has
localized distributions at $\alpha=1$, such as $\delta(\alpha-1)$. The
linearity assumption is backed by explicit calculations in various
non-perturbative
models~\cite{Basham:1978zq,Dokshitzer:1995qm,Korchemsky:1999kt,Belitsky:2001ij,
  Schindler:2023cww}, that we have verified with an explicit
calculation in the dispersive approach of
Ref.~\cite{Dokshitzer:1995qm}.

The power of the light-ray OPE~\eqref{eq:light-ray_OPE_int} is that it allows us to gain an understanding of the collinear parametrization beyond first the power correction as well as of their quantum evolution giving rise to the quantum scaling violation.

From~\eqref{eq:light-ray_OPE_int}, we can qualitatively infer the reason why our prediction for the evolution of the leading power correction does not describe very well the EEC data while providing a good description for the higher-point projective correlators (see FIG.~\ref{fig:ee_plot}).
In the EEC case, the leading power correction consists of $J=2$ operators, whose anomalous dimension matrix $\gamma_{\tau=2}^{[J=2]}$ contains one vanishing eigenvalue. This explains the reason why the leading power correction has a very moderate scale evolution. On the other hand, the subleading contributions of order $\mathcal{O}(\Lambda_{\rm QCD}^{(\alpha\approx 2)})$ are sensitive to the $J=1$ pole in the anomalous dimensions, thus they are expected to have larger scaling evolution effects and require BFKL resummation when the label $J$ is very close to $1$~\cite{Lipatov:1996ts,Brower:2006ea,Jaroszewicz:1982gr,Kotikov:2000pm,Kotikov:2002ab}.
However, for $N\geq 3$ the evolution of the next few subleading-power corrections is milder than that in the leading power correction because the eigenvalues of $\gamma_{\tau=2}^{[J>2]}$ are positive and monotonically increasing with $J$. In these cases, the subtleties due to the $J=1$ pole are delayed to higher power corrections.

%%%%%%%%%%%%%%%%%%%%%%%%%%%%%%%%%%%%%%%%%%%%%%%%%%%%%%%
\subsection{Leading Logarithmic Approximation}

In this section, we discuss the calculation of the leading power correction quantities $\text{ENC}_{1,\Psi_q}^{\rm N.P.}(K_\perp, Q)$ for various states $\{ |\Psi_q\rangle \}$ and energy scales $\{Q\}$ at Leading-Logarithmic (LL) accuracy\footnote{When presenting the light-ray OPE method, Eq.~\eqref{eq:EEC_power_corr} is the result of a simplified calculation at leading-logarithmic accuracy. On the other hand, the factorization formula \eqref{eq:factorization_formula} extends to all logarithmic orders where the variable $x$ inside the jet function plays a very important role. This requires including derivatives of light-ray operators in the light-ray OPE. For technical details, one can follow Ref.~\cite{Chen:2023zzh} to generalize \eqref{eq:EEC_power_corr} to all logarithmic orders, which will be equivalent to the factorization formula \eqref{eq:factorization_formula}.}.
The starting point is Eq.~\eqref{eq:EEC_power_corr} and we set the factorization scale $\mu=K_\perp$
\begin{equation}
  \text{ENC}_1^{\rm N.P.}(K_\perp, Q) = \Lambda_{\rm QCD}
    \, \vec{D}_N \left(1, \frac{\Lambda_{\rm QCD}^2}{K_\perp^2} \right) \cdot 
   \left[\frac{\langle  \vec{\mathbb{O}}_{\tau = 2}^{[J=N]}(n;\mu) \rangle_{\Psi_q}}{{(4\pi)^{-1}\sigma_{\Psi_q}   Q^{N-1}}} \left( \frac{Q^2}{\mu^2}\right)\right]\Bigg|_{\mu = K_\perp}\,.\notag
 \end{equation}
 Assuming that the state $|\Psi_q\rangle$ is physical and hence it does not depend on the factorization scale $\mu$, the light-ray operator renormalization equation~\eqref{eq:light-ray_RG} gives the corresponding evolution equation for the matrix element
 \begin{equation}
   \mu \frac{d}{d\mu} \langle \vec{\mathbb{O}}_{\tau = 2}^{[J]}(n;\mu)\rangle_{\Psi_q} = \gamma_{\tau=2}^{[J]}(\mu)\cdot \langle \vec{\mathbb{O}}_{\tau = 2}^{[J]}(n;\mu)\rangle_{\Psi_q}\,,
 \end{equation}
 whose solution can be generated by the evolution operator in \eqref{eq:RG_kernel} \begin{equation}
 \langle \vec{\mathbb{O}}_{\tau = 2}^{[J]}(n;\mu)\rangle_{\Psi_q} = U_N(\mu, Q) \cdot  \langle \vec{\mathbb{O}}_{\tau = 2}^{[J]}(n;Q)\rangle_{\Psi_q}\,.
\end{equation}
This leads to the main result in the Letter \eqref{eq:evolution}. In the LL approximation, the evolution matrix is determined by the leading-order anomalous dimension matrix $\gamma_{\tau = 2}^{[J]}(\mu) = \frac{\alpha_s(\mu)}{4 \pi} \gamma_{\tau =2}^{[J],(0)} +\mathcal{O}(\alpha_s^2)$ and $\gamma_{\tau =2}^{[J],(0)}$ is~\cite{Gross:1974cs,Gross:1973ju}
\begin{equation}
\gamma_{\tau =2}^{[J],(0)}=\begin{pmatrix}
 2C_F(4H_J-\frac{2}{J(J+1)}-3)&
 -T_F \frac{4(J^2+J+2)}{J(J+1)(J+2)}\\
-C_F\frac{4(J^2+J+2)}{(J-1)J(J+1)}&
8C_A(H_J-\frac{2(J^2+J+1)}{(J-1)J(J+1)(J+2)})-2\beta_0\\
\end{pmatrix}\,,
\end{equation}
where $H_J=\sum_{n=1}^J \frac{1}{n}$ is the harmonic number, $\beta_0=\frac{11}{3}C_A-\frac{4}{3}n_f T_F$ is the one-loop QCD beta function. The path ordering exponential in \eqref{eq:RG_kernel} simplifies at LL and gives
\begin{equation}\label{eq:LL_kernel}
U_N^{\rm LL}(K_\perp,Q) = \left[\frac{\alpha_s(Q)}{\alpha_s(K_\perp)}\right]^{\gamma_{\tau=2}^{[N],(0)}/(2\beta_0) }\,.
\end{equation}

The matrix element $\langle \vec{\mathbb{O}}_{\tau = 2}^{[J]}(n;Q)\rangle_{\Psi_q}$ can be approximated by the perturbative result because we work under the assumption of the hierarchy $Q \gg \Lambda_{\rm QCD}$. Further hadronization corrections to this matrix element are not enhanced in the collinear limit and hence can be neglected. In the LL approximation, we only need the leading-order contribution in the process-dependent quantity $\frac{\langle \vec{\mathbb{O}}_{\tau = 2}^{[J=N]}(n_k;Q) \rangle_{\Psi_q} }{(4\pi)^{-1}\sigma_{\Psi_q} Q^{N-1}}$.
This is $(2^{2-N},0)^T+\mathcal{O}(\alpha_s)$ for $\gamma^*\to q\bar{q}$ and $(0, 2^{2-N})^T+\mathcal{O}(\alpha_s)$ for $h\to gg$, from which we can extract out the quark jet and gluon jet leading non-perturbative correction functions, respectively.
We now denote all parts in the r.h.s. of Eq.~\eqref{eq:evolution} independent of the state $|\Psi_q\rangle$ (when $Q$ is fixed) as $\vec{F}^{[N]}_1(K_\perp, Q)$:
\begin{equation} \label{eq:FN_def}
  \vec{F}^{[N]}_1(K_\perp, Q)\equiv \Lambda_{\rm QCD} \vec{D}_N \left(1, \frac{\Lambda_{\rm QCD}^2}{K_\perp^2} \right) \! \cdot U_N(K_\perp, Q)\,.
\end{equation}
This vector contains the information of the quark and gluon leading non-perturbative functions as well as the perturbative scale evolution. From the definition \eqref{eq:FN_def}, we can find that $\vec{F}^{[N]}_1$ evolves as follows between two scales $Q$ and $Q_0$
\begin{equation} \label{eq:FN_property}
  \vec{F}^{[N]}_1(K_\perp, Q)=\vec{F}^{[N]}_1(K_\perp, Q_0)\cdot U_N(Q_0, Q)\,.
\end{equation}
The LL predictions for $\gamma^*\to qq$ and $h\to gg$ are then
\begin{align}
 \text{ENC}_{1,\gamma^*\to q\bar{q}}^{\rm N.P.}(K_\perp, Q)&= \vec{F}^{[N]}_1(K_\perp, Q)  \cdot  
  \begin{pmatrix}
  2^{2-N}\\
  0
  \end{pmatrix}\,,\\
   \text{ENC}_{1,h\to gg}^{\rm N.P.}(K_\perp, Q)&= \vec{F}^{[N]}_1(K_\perp, Q)  \cdot  
  \begin{pmatrix}
  0\\
  2^{2-N}
  \end{pmatrix}\,,
\end{align}
from which we can deduce the following LL relation
\begin{equation}
\begin{pmatrix}
  \text{ENC}_{1,\gamma^*\to q\bar{q}}^{\rm N.P.}(K_\perp,Q) \\
    \text{ENC}_{1,h\to gg }^{\rm N.P.}(K_\perp,Q)
\end{pmatrix}_{\rm LL}^T
=
\begin{pmatrix}
  \text{ENC}_{1, \gamma^*\to q\bar{q}}^{\rm N.P.}(K_\perp,Q_0) \\
    \text{ENC}_{1, h\to gg}^{\rm N.P.}(K_\perp,Q_0)
\end{pmatrix}_{\rm LL}^T
\cdot U^{\rm LL}_N(Q_0, Q)\,.
\end{equation}
Similarly, we obtain a similar relation for quark and gluon jets in the $pp\to Z+$jet case
\begin{equation}
\begin{pmatrix}
  \text{ENC}_{1,pp\to Zq}^{\rm N.P.}(K_\perp, E_J)\\
    \text{ENC}_{1,pp\to Zg }^{\rm N.P.}(K_\perp, E_J)
\end{pmatrix}_{\rm LL}^T
=
\begin{pmatrix}
  \text{ENC}_{1, pp\to Zq}^{\rm N.P.}(K_\perp, E_{J,0})\\
    \text{ENC}_{1, pp\to Zg}^{\rm N.P.}(K_\perp, E_{J,0})
\end{pmatrix}_{\rm LL}^T
\cdot U^{\rm LL}_N(E_{J,0}, E_J)\,,
\end{equation}
where $E_J$ is the jet energy and we have used $K_\perp = R_L E_J$ as
in the Letter. The validation of these relations are described in the
section {\it Monte Carlo validation}.

%%%%%%%%%%%%%%%%%%%%%%%%%%%%%%%%%%%%%%%%%%%%%%%%%%%%%%%
\subsection{Different strategies to predict scaling violation and
  subleading power corrections}
We now discuss the extraction of the non-perturbative boundary
conditions to Eq.~\eqref{eq:F-LL} that fix the two entries of the
vector $\vec{D}$.
We can take two strategies, discussed in the following.

The first strategy is to extract the two non-perturbative entries of
$\vec{D}$ at a fixed energy $Q=Q_0$, using both quark and gluon data
(for instance via the $\gamma^*\to q\bar{q}$ and $h\to gg$
decays). This is the strategy adopted in the Letter, and resulting in
the predictions shown in FIG.~\ref{fig:ee_plot} and
FIG.~\ref{fig:pp_plot}. The advantage of this strategy is that it
extracts simultaneously quark and gluon distributions. Moreover, this
allows us to target more precisely the leading power correction
${\cal O}(\Lambda_{\rm QCD}/Q)$ provided the scale $K_\perp$ is
sufficiently large.

\begin{figure}[htbp]
\begin{center}
\includegraphics[width=0.5\textwidth]{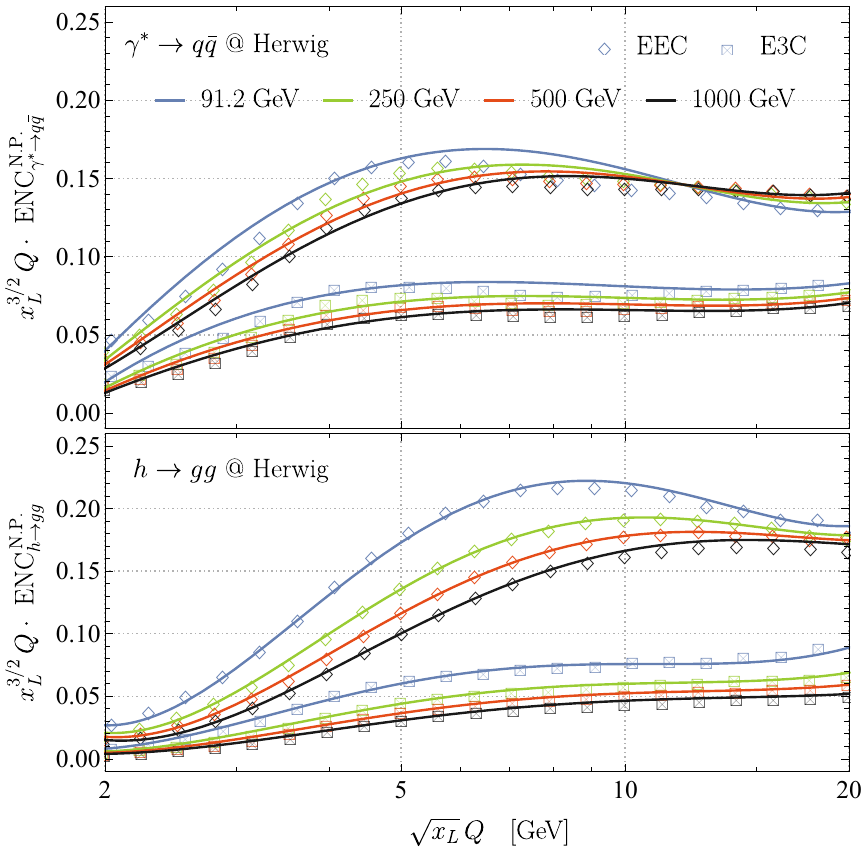}
\caption{Comparison of approximate relation \eqref{eq:approx_relation}
  and Monte Carlo predictions for the quantum scaling violation for
  $N$-point projected correlators in $\gamma^*\to q\bar{q}$ and
  $h\to gg$.  The $250\, \mathrm{GeV}$ curves (green) and the
  $500\, \mathrm{GeV}$ curves (red) are extracted from Monte Carlo
  data, while the $91.2\, \mathrm{GeV}$ curves (blue) and the
  $1000\, \mathrm{GeV}$ curves (black) are predictions from
  \eqref{eq:approx_relation}.}
\label{fig:opt1}
\end{center}
\end{figure}

The second strategy is instead to use either only quark or only gluon
data at \textit{two} different energies $Q_1$ and $Q_2$. One advantage
of using two different energy scales is that this strategy partly
captures subleading power corrections, and hence it is expected to
work better at lower energies as well as in the EEC case, for which we
have found that higher power corrections are enhanced
(cf. FIG.~\ref{fig:ee_plot}).
On the other hand, one disadvantage is that it cannot be used to
extract, simultaneously, the quark and gluon distributions since we
only have two degrees of freedom.
To show how to implement this, we can diagonalize the LL kernel
\eqref{eq:LL_kernel} using the right eigenvectors of
$\gamma_{\tau=2}^{[N],(0)}$
\begin{equation}
\gamma_{\tau=2}^{[N],(0)}\cdot \vec{v}_i^{[N]} = \lambda_i^{[N]} \vec{v}_i^{[N]}\,,
\end{equation}
where $\lambda_{i=1,2}^{[N]}$ is the eigenvalue corresponding to $i$-th right eigenvector $\vec{v}_i^{[N]}$.
As a result, the LL approximation for Eq. \eqref{eq:FN_def} becomes
\begin{equation}\label{eq:F-LL}
\vec{F}^{[N]}_1(K_\perp, Q)\cdot \vec{v}_i^{[N]}= \left[\frac{\alpha_s(Q)}{\alpha_s(K_\perp)}\right]^{\lambda_i^{[N]}/(2\beta_0) }
\left[ \Lambda_{\rm QCD} \vec{D}_N \left(1, \frac{\Lambda_{\rm QCD}^2}{K_\perp^2} \right) \cdot \vec{v}_i^{[N]} \right]\,.
\end{equation}
From Eq.~\eqref{eq:F-LL} we can establish the relation
\begin{equation}
\label{eq:relation_2}
\vec{F}_1^{[N]}(K_\perp,Q) = \frac{R_N(Q,Q_2) }{R_N(Q_1,Q_2)} \vec{F}_1^{[N]}(K_\perp, Q_1)+ \frac{R_N(Q_1,Q)}{R_N(Q_1,Q_2)} \vec{F}_1^{[N]}(K_\perp, Q_2)\,,
%\vec{F}_1^{[N]}(Q) = \sum_{i,j=1}^2 \epsilon_{ij} \frac{R(Q,Q_j)}{R(Q_i,Q_j)} \vec{F}_1^{[N]}(Q_i)\,,
\end{equation}
%where $\epsilon_{ij}$ is 2d Levi-Civita symbol with $\epsilon_{12}=-\epsilon_{21}=1$ and $\epsilon_{ii}=0$. 
where $R_N$ is defined as the determinant
\begin{equation}
R_N(Q_1,Q_2) = \det
\begin{pmatrix}
\alpha_s(Q_1)^{\lambda_1^{[N]}/(2\beta_0)} &\alpha_s(Q_1)^{\lambda_2^{[N]}/(2\beta_0)}\\
\alpha_s(Q_2)^{\lambda_1^{[N]}/(2\beta_0)} &\alpha_s(Q_2)^{\lambda_2^{[N]}/(2\beta_0)}
\end{pmatrix}\,.
\end{equation}
It turns out that ratios of $R$ functions in Eq.~\eqref{eq:relation_2}
is numerically not very sensitive to the value of anomalous dimensions
or $N$.  Therefore, such a relation may numerically, and
approximately, capture terms beyond the leading power correction
because we expect that the next few contributions are described by the
anomalous dimensions at nearby $J<N$.
This leads to the approximate relation
\begin{equation}\label{eq:approx_relation}
%\frac{\text{ENC}^{\rm N.P.}(K_\perp, Q) }{Q^2 K_\perp^{-3}}
%\approx \sum_{i,j} \epsilon_{ij} \frac{R(Q,Q_j)}{R(Q_i,Q_j)} \frac{\text{ENC}^{\rm N.P.}(K_\perp, Q_i) }{Q_i^2 K_\perp^{-3}}\,,
x_L^{3/2} Q\,\text{ENC}_{\Psi_q}^{\rm N.P.}(K_\perp, Q) 
\approx 
\frac{R_N(Q,Q_2) }{R_N(Q_1,Q_2)}  x_L^{3/2} Q_1\,\text{ENC}_{\Psi_q}^{\rm N.P.}(K_\perp, Q_1) 
+ \frac{R_N(Q_1,Q)}{R_N(Q_1,Q_2)} x_L^{3/2} Q_2\, \text{ENC}_{\Psi_q}^{\rm N.P.}(K_\perp, Q_2)\,,
\end{equation}
which provides reliable predictions over a wider range of $K_\perp$
near the transition region.
We show the results of this procedure applied separately to quark and
gluon final states in FIG.~\ref{fig:opt1}. Here we choose
$Q_1=250\, \mathrm{GeV}$, $Q_2=500\, \mathrm{GeV}$, and generate Monte
Carlo data for both $\gamma^*\to q\bar{q}$ and $h\to gg$ processes as
input, and make predictions at $Q=\{91.2, 1000\}$\GeV. We see that
this procedure leads to a very good description of the data in the
region $\sqrt{x_L} Q \lesssim 15 \, \mathrm{GeV}$ where leading power
correction predictions fail in EEC, indicating that
\eqref{eq:approx_relation} could approximately capture also the
evolution of the subleading power corrections.

An important aspect of FIG.~\ref{fig:opt1} is that quark and gluon
distributions are extracted separately, that is we extract four
non-perturbative distributions rather than the usual two used in
Eq.~\eqref{eq:LP_ee_higgs}.
This clearly makes it impossible to describe quark and gluon data with
a single set of non-perturbative functions of $K_\perp$.
A rigorous way of dealing with this problem would be to consider the
consistent inclusion of subleading power corrections in the OPE.

\subsection{Impact of quantum scaling violation on ratios of energy correlators}
\label{app:enc-ratios}
In the Letter we have shown that the dominant classical scaling of the
power corrections to energy correlators receives a correction due to
quantum scaling violation. Although the latter is numerically
subdominant at the level of the individual correlator, its effect is
instrumental when one considers ratios of
correlators~\cite{Chen:2020vvp} such as $\text{E3C}/\text{EEC}$. The
measurement of ratios of this type has been used for recent
extractions of the strong coupling constant by the CMS
collaboration~\cite{CMS:2024mlf}.

In this section we discuss the impact of quantum scaling violation on
the ratio of energy correlators $\text{ENC}/\text{EEC}$. Explicitly,
we define the non-perturbative correction to such ratios as
\begin{equation}
  R_{\Psi}^{\text{N.P},\,N/2}(x_L,Q) \equiv
  \left[\frac{\text{ENC}_{\Psi}(x_L,Q)}{\text{EEC}_{\Psi}(x_L,Q)}\right]^{\rm Hadron}-  \left[\frac{\text{ENC}_{\Psi}(x_L,Q)}{\text{EEC}_{\Psi}(x_L,Q)}\right]^{\rm Parton}\,,
\end{equation}
where as usual $\Psi = \{\gamma^*\to q\bar{q},h\to gg\}$ denotes the
specific final state in which the correlator is measured.

\begin{figure}[htbp]
\begin{center}
  \includegraphics[width=0.49\textwidth]{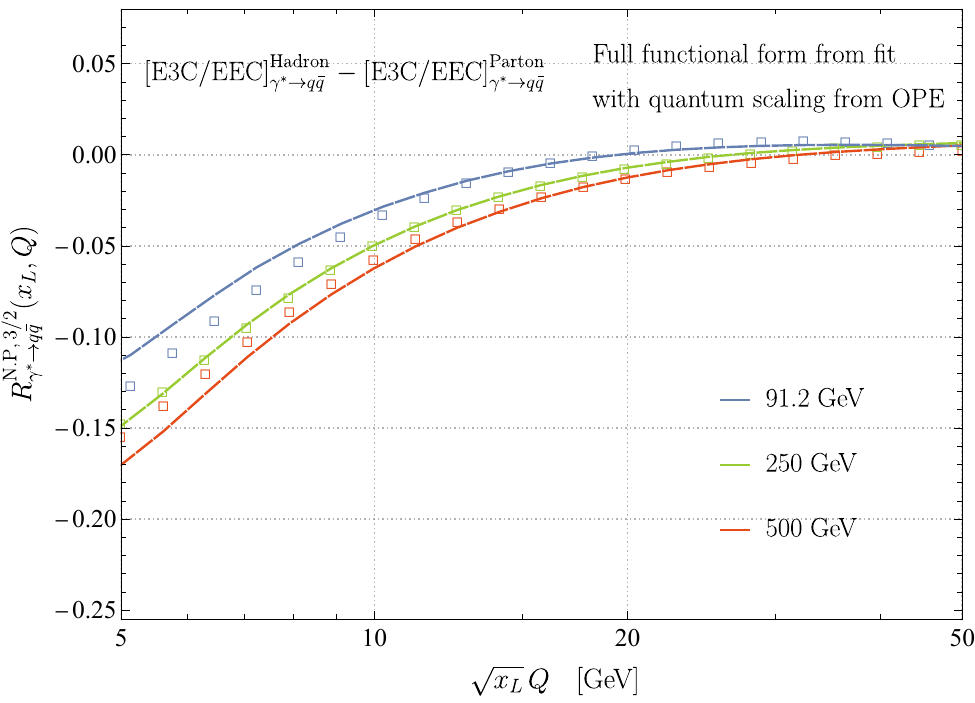}
  \includegraphics[width=0.49\textwidth]{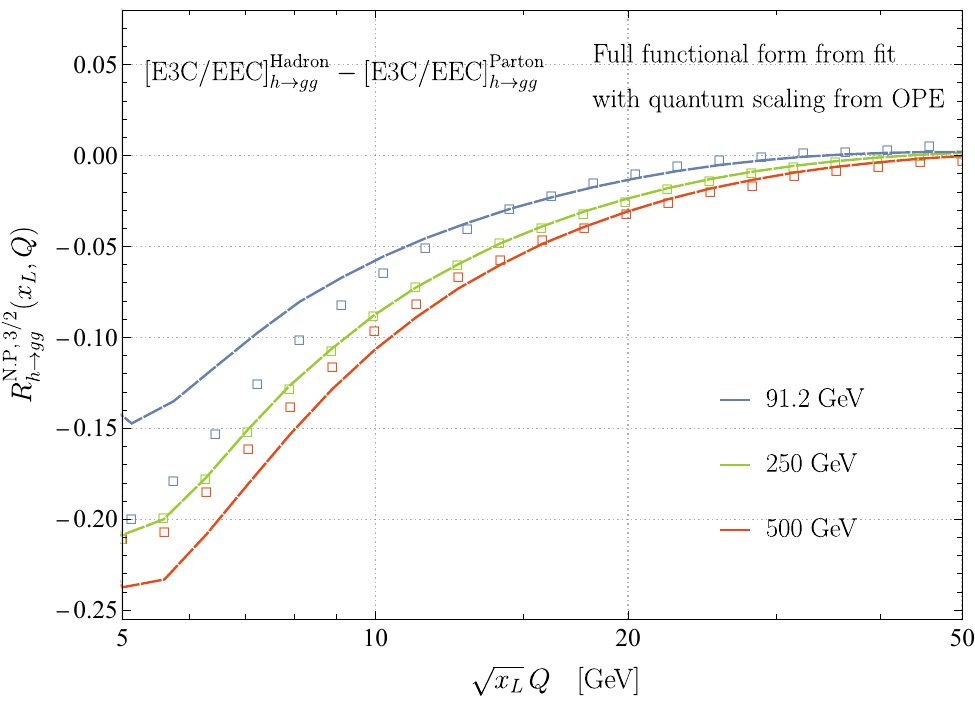}
  \caption{Leading power correction for the ratio
    $\text{E3C}/\text{EEC}$ measured in $\gamma^*\to q\bar{q}$ (left)
    and $h\to gg$ (right). The solid lines represent the predictions
    from the quantum scaling violation derived in the Letter, where
    the non-perturbative function of $\sqrt{x_L}Q$ is obtained from a
    fit of Monte Carlo data $Q=250$ GeV and then evolved using the
    evolution equation given in the Letter. The squares are the result of
    Monte Carlo simulations.}
\label{fig:ratio}
\end{center}
\end{figure}

We explicitly consider the ratio $\text{E3C}/\text{EEC}$, whose power
corrections are shown in Fig.~\ref{fig:ratio} for
$\gamma^*\to q\bar{q}$ (left plot) and $h\to gg$ (right plot). The
dominant effect due to classical scaling largely cancels at the level
of the ratio, hence enhancing the effect of quantum scaling
violation. In the figures, the latter effect is responsible for the
spread between data at different collision energies, which is not
predicted by the classical scaling (i.e. all curves would be otherwise
completely overlapping). Therefore the effect of quantum scaling is
substantial at the level of the ratios, hence playing a significant
role in experimental extractions of the strong coupling constant.

As done in the main text, we compare our prediction for the leading
power correction to a Monte Carlo simulation obtained with Herwig 7.2,
for both quark and gluon processes. The Monte Carlo data are
represented by the squares, while the lines represent our
prediction. To obtain that, we fit the non-perturbative $\sqrt{x_L}Q$
dependence at the reference energy $Q=250$ GeV, and then evolve in $Q$
using the results of the Letter. As it can be seen from
Fig.~\ref{fig:ratio} the evolution of the quantum scaling leads to a
rather good description of the Monte Carlo simulation. We note that
the agreement deteriorates for $\sqrt{x_L}Q\lesssim 10$ GeV, due to
the fact that the $\text{EEC}$ suffers from soft-gluon effects at the
level of sub-leading power corrections. Nevertheless, it can be also
seen that soft gluon effects are suppressed for the ratio compared to
$\text{EEC}$ itself, as reflected in Fig.~\ref{fig:ratio}. It would be
interesting to quantify how the quantum scaling affects extractions of
the strong coupling constant from the $\text{E3C}/\text{EEC}$ ratio,
which we leave for future work.